\documentstyle[prd,aps,psfig,eqsecnum]{revtex}
\begin{document}

\twocolumn

\def\vm{v_{\rm max}}
\def\beq{\begin{equation}}
\def\eeq{\end{equation}}
\def\la{\langle}
\def\ra{\rangle}
\def\k{\kappa}
\def\b{\beta}
\def\d{\delta}
\def\o{\omega}
\def\gsim{\; \raisebox{-.8ex}{$\stackrel{\textstyle >}{\sim}$}\;}
\def\lsim{\; \raisebox{-.8ex}{$\stackrel{\textstyle <}{\sim}$}\;}
\def\gtrsim{\gsim}
\def\lessim{\lsim}
\def\phi{\varphi}

\title{Hawking radiation on a falling lattice} 
 
\author{Ted Jacobson$^{a,b}$\thanks{jacobson@physics.umd.edu} 
and  David Mattingly$^a$\thanks{davemm@physics.umd.edu}}
\address{$^a$Department of Physics, University of Maryland,
College Park, MD 20742-4111, USA}
\address{$^b$Institute for Theoretical Physics, UCSB, Santa Barbara, CA 93106}
\maketitle
\begin{abstract}
Scalar field theory on a lattice falling freely into a 1+1 dimensional
black hole is studied using both WKB and numerical approaches. The outgoing
modes are shown to arise from incoming modes by a process analogous to
a Bloch oscillation, with an admixture of negative frequency modes 
corresponding to the Hawking radiation. Numerical calculations show that
the Hawking effect is reproduced to within 0.5\% on a lattice
whose proper spacing where the wavepacket turns around 
at the horizon is $\sim0.08$ in units where the surface gravity is 1. 
\end{abstract}
\pacs{04.70.Dy, 04.60.Nc}
  
\section{INTRODUCTION}

If nothing can emerge from a black hole then where does the Hawking
radiation come from? 
Relativistic field theory tells us that
it comes from
a trans-Planckian reservoir of outgoing modes at the
horizon\cite{TJultra}.  
The existence of such a trans-Planckian reservoir is doubtful on general
physical grounds. It is the cause of the divergences in quantum
field theory which are expected to be removed in a quantum theory of
gravity, and in particular it would seem to imply an infinite black
hole entropy unless canceled by an infinite negative ``bare entropy".

String theory has produced an account of the 
Hawking radiation and black hole entropy which requires no 
trans-Planckian reservoir,
however in that picture instead of a black hole one has 
a stringy object in flat spacetime 
that arises from a black hole as the string coupling
constant tends to zero\cite{Peet}. 
Thus gravitational redshift plays no role
in the string calculations and nothing is learned about the origin of
the outgoing modes in a black hole spacetime.\footnote{The fact 
that redshifting plays no role in producing the stringy Hawking 
radiation is presumably tied to the fact that the string calculations
apply in the ``near-extremal" limit that the surface gravity tends to zero.}
If we could follow the theory as the string coupling becomes
strong we would presumably learn how it is that string theory produces 
the outgoing black hole modes.

Absent a quantum gravity description of the Hawking process, 
Unruh invented a fluid analog of a black hole\cite{Unruh81}
in which an inhomogeneous flow exceeds the long wavelength 
speed of sound creating a sonic horizon 
(see also \cite{Moncrief,Visser}).
Quantizing the sound field Unruh argued that the horizon would radiate
thermal phonons, in analogy with the Hawking effect, at a temperature
given by $\hbar/2\pi$ times the gradient of the flow 
velocity at the horizon. 
Since the sound field has an atomic cutoff, and the
physics of the fluid is completely understood in principle, the fluid
model may provide a source of insight into the Hawking process.

The atomic cutoff produces dispersion of sound waves leading to
subsonic propagation at high frequencies. It turns out that this
dispersion is all that is needed to obviate the need for a 
trans-Bohrian reservoir. In the linearized theory,
an outgoing short wavelength mode
can be dragged in by the fluid flow and redshifted enough
so that its increased group velocity overcomes the flow and it
escapes to infinity. 
Investigations of a number of linear field 
theory 
models\cite{Unruh95,BMPS,CJ,Corleyanalytic,HimemotoTanaka,SaidaSakagami} 
and lattice models\cite{CJlattice} 
incorporating such high frequency dispersion 
have demonstrated this mechanism of ``mode conversion"  
and have shown that the Hawking radiation is recovered.\footnote{ 
Dispersion can also produce superluminal
propagation, which allows outgoing modes to emerge from behind the horizon.
The superluminal case does not seem very healthy from a fundamental
point of view, though it is surely relevant in some condensed matter
analogues of black hole horizons\cite{JacoVolo,KopnVolo,Volotorus} 
and is capable of producing the
usual Hawking spectrum\cite{Corsuper}.} The dispersion in these models 
is not locally Lorentz-invariant since a frame is
picked out in which to specify which frequencies are ``high".
In the fluid model this is the local rest frame of the fluid. 
A real black hole also defines a preferred frame
but it does so in a non-local way. Perhaps quantum gravity
produces a dispersion effect related to this non-local notion of a
preferred frame, or perhaps microphysics is just not locally Lorentz
invariant. In any case, it seems worthwhile to understand the mechanism
of the outgoing modes and Hawking radiation in these models for the
hints it may provide about a correct quantum gravity account.

In \cite{CJlattice} a falling lattice model was introduced 
to address the ``stationarity puzzle"\cite{TJOrigin,CJ}: 
how can a low frequency outgoing
mode arise from a high frequency ingoing mode
when it propagates only in a stationary background spacetime  
and hence has a conserved Killing frequency? 
In the continuum based dispersive models
there is no satisfactory resolution of this puzzle.
As the outgoing modes are traced backwards in time their
incoming progenitors are squeezed into the short distance
regime of the model which is not physically sensible.

A lattice provides a simple model for imposing a physically 
sensible short distance cutoff. It might seem at first that
the most natural choice would be to preserve the time translation
symmetry of the spacetime with a static lattice whose points
follow accelerated worldlines. On a static lattice, however, the
Killing frequency is conserved, so outgoing modes arise from
ingoing modes with the same frequency, and there is no Hawking
radiation. The in-vacuum therefore evolves to a singular state at the
horizon. If the lattice points are instead freely falling,  
a discrete remnant of time translation symmetry can still be preserved.
On such a lattice there is
still a frequency conservation law, and the unsatisfactory 
resolution of the stationarity puzzle is that the lattice spacing 
in this case goes to zero at infinity (if the 
lattice points are asymptotically at rest)\cite{CJlattice}.

If we insist that the lattice spacing asymptotically
approaches a fixed constant {\it and} that the lattice points are at rest
at infinity, we find a fascinating resolution of the stationarity
puzzle\cite{CJlattice}:  the lattice cannot have even a discrete time
translation symmetry, since there is a gradual spreading
of the lattice points as they fall toward the horizon. The timescale
of this spreading is of order $1/\k$ where $\k$ is the surface
gravity. This time dependence of the
lattice is invisible to long wavelength modes which sense
only the stationary background metric of the black hole, but it is
quite apparent to modes with wavelengths of order the lattice spacing.
On such a lattice the long wavelength outgoing modes 
come from short wavelength ingoing modes which start out with a high
(lattice scale) Killing frequency which is exponentially redshifted
as the wavepacket propagates to the horizon and turns around. 

This behavior of wavepackets in the (time-dependent)
falling lattice model was determined
in \cite{CJlattice} with the help of the WKB approximation in which 
the frequency is used as a Hamiltonian to solve for the wavepacket 
trajectory. It was also argued that, since the 
time dependence of the lattice is adiabatic for those modes with
wavelength short enough to know they are on a lattice, 
these modes will remain unexcited on their way to the horizon
and the ground state condition for the Hawking effect will be met.
However, since the wavelength is of order the lattice spacing and 
the wavepacket can vary wildly on the lattice near the horizon, 
it is not obvious that the WKB approximation is reliable nor that 
the adiabatic argument really applies. 

The primary motivation for
the work reported here was to check the WKB and adiabatic
assumptions by carrying out the exact calculation numerically 
using the lattice wave equation. We found that indeed the
field behaves this way, and the Hawking radiation is recovered 
to within half a percent for a lattice spacing
$\d=0.002/\k$ which corresponds to a proper spacing 
$\sim 0.08/\k$ where the wavepacket turns around at the horizon. 
We also studied the deviations
from the Hawking effect that arise as $\kappa\d$ is increased, and
our simulations reveal an interesting picture of how the wavepackets
turn around at the horizon.

The precise choice of worldlines for the lattice points should not
be important to the leading order Hawking effect as long as the ground
state evolves adiabatically in the lattice theory. Our numerical results
are consistent with this expectation, in that the particular worldlines
depend on the lattice spacing yet the results converge to the continuum 
as the lattice spacing is decreased. Moreover it was shown 
recently\cite{HimemotoTanaka},
in a continuum model with high frequency dispersion, that when the
preferred frame is changed from the free-fall frame of the black hole 
to that of a conformally related metric there is no leading order change
in the Hawking radiation
unless the acceleration of the preferred frame is very large. 
Presumably all that matters is that the preferred 
world lines flow smoothly across the horizon. From a fundamental point of 
view, if there really is a cutoff in some preferred frame, it is plausible 
that this frame would coincide with the cosmic rest frame and would fall 
across the event horizons of any black holes that form by collapse.

The remainder of this paper is organized as follows. In 
Section \ref{flm}
the falling lattice model is set up, in Section \ref{wkb} some results
of the WKB method are shown, illustrating the frequency shift and the
mode conversion at the horizon, and the role of the negative frequency
branch is explained. In 
Section \ref{numerical} the results of the full numerical
evolution are presented, and Section \ref{discussion} contains a discussion
or the results and directions for further work. The Appendix describes the
finite differencing of the time derivates used in the numerical calculations.

We use units with $\hbar=c=\kappa=1$, where $\k$ is the
surface gravity.

\section{FALLING LATTICE MODEL}
\label{flm}
In this section we set up the field theory on a lattice falling
into a black hole in two spacetime dimensions.

\subsection{Black hole spacetime} 
We assume the spacetime metric is static, so\cite{Painleve}
coordinates can be chosen (at least locally)
such that the line element takes the form
\beq
ds^2 = dt^2 - \bigl(dx - v(x) dt\bigr)^2.
\label{ds2}
\eeq
In these coordinates the 
Killing vector is given by
\beq
\chi=\partial_t,
\label{kv}
\eeq
with squared norm
\beq
\chi^2 = 1- v^2(x).
\eeq
For $v(x)$ we choose 
\beq
v(x)=-\frac{v_{\rm max}}{\cosh(\b x)},
\label{v}
\eeq
where $v_{\rm max}$ and $\b$ are positive constants. 
As $|x|\rightarrow +\infty$, $v(x)$ vanishes, so the line element 
becomes that of Minkowski spacetime.  
Provided $ v_{\rm max} >1$ there is an ergoregion centered on $x=0$
in which the Killing vector is spacelike, and the boundaries of this
region are black and white hole horizons at
\beq
x_H=\pm \b^{-1}\cosh^{-1}( v_{\rm max} ). 
\label{xH}
\eeq
The surface gravity of the horizons is given by 
\beq
\kappa=v'(x_H)=\b\sqrt{1- v_{\rm max}^{-2}}.
\eeq
The surface gravity sets the length scale for our spacetime, and
we will keep it fixed. Hence it will be useful to employ units
for which $\kappa=1$. 
 
We now introduce a freely falling ``Gaussian" coordinate $z$
which will be discretized to define a falling lattice.
The worldlines with $dx=v(x)dt$ are geodesics of the metric
(\ref{ds2}) with proper time $t$, at rest as $|x|\rightarrow \infty$,
and they are orthogonal to the surfaces of constant $t$. 
On these curves the quantity $t-\int dx/v$ is constant,
so if $z$ is constant on these curves we must have 
$W(z(x,t))=t-\int dx/v$ for some function $W$. 
We choose this function
so that $z=x$ at $t=0$, which implies that 
$W(z)=-\int^z dx/v=\sinh(\b z)/\b\vm$, hence
\beq
\sinh(\b z)= \sinh(\b x) + \b\vm t .
\eeq
In terms of $z$, the line element (\ref{ds2}) takes the form
\beq
ds^2 = dt^2 - a^2(z,t)\, dz^2,
\label{ds2z}
\eeq
with
\beq
a(z,t)= \frac{\cosh(\b z)}{\sqrt{1+(\sinh(\b z)-\b\vm t)^2}}.
\eeq
We call $a(z,t)$ the ``scale factor", even though it depends
on $z$ as well as $t$. It can also be expressed as
\beq
a(z,t)= \frac{v(x(z,t))}{v(z)}.
\eeq
In $(t,z)$ coordinates
the Killing vector (\ref{kv}) is given by 
\beq
\chi=\partial_t - v(z) \partial_z=
\partial_t - \frac{v(x(z,t))}{a(z,t)} \partial_z.
\label{kvz}
\eeq

The scale factor is unity at $t=0$, and outside the black hole it
grows towards the future. It is important to know how rapidly the 
scale factor is changing in time, since in the lattice theory that 
time dependence is not merely a coordinate effect and can therefore
excite the quantum field. At the horizon the scale factor takes 
the value 
\beq
a(z_H,t)=\sqrt{1+2\k t + \k^2t^2/(1-\vm^{-2})},
\label{aH}
\eeq
so $a(z_H,t)\sim \k t$ for $\k t\gtrsim 1$. Thus the fractional
rate of change $\partial_t a/a$ is of order $\k$. In general,
we have
\beq
\partial_t a/a = 
\frac{\b\vm\sinh(\b x)}{\cosh^2(\b x)}
\eeq
whose maximum outside the horizon ($\cosh(\b x)>\vm$)
is $\k$ (at the horizon) if $\vm\ge\sqrt{2}$ and
$\k\vm^2/2\sqrt{\vm^2-1}$ if $\vm\le\sqrt{2}$.
Thus $|\partial_t a/a|\lessim O(\k)$ everywhere
outside the horizon
as long as $\vm>1$ is not too close to unity.

\subsection{Lattice}
\begin{figure}[hbt]
\centerline{
\psfig{figure=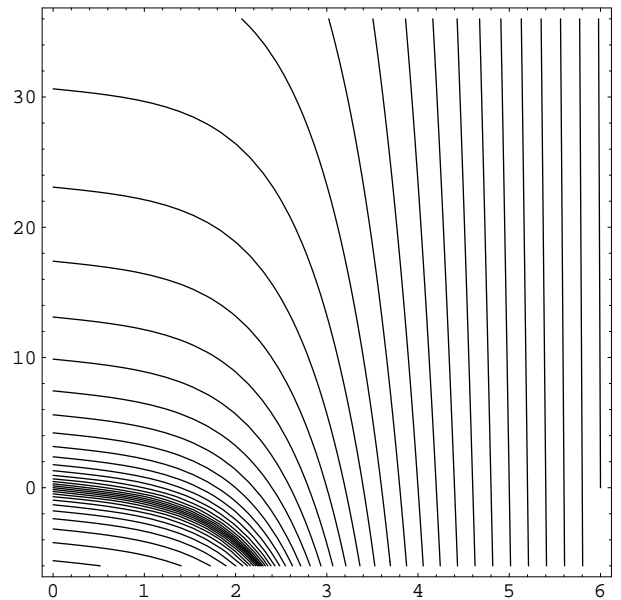,angle=0,height=8cm}}
\vskip 2mm
\caption{\small Every 50th lattice point, $t$ vs. $x$, 
for the case $\d=0.004$ and 
$v_{\rm max}=\protect\sqrt{2}$. 
In the black region of the plot
the points are too close to resolve. The horizon is at $x\simeq 0.6$.} 
\label{latticefig}
\end{figure}

The lattice is defined by discretizing the $z$ coordinate
with spacing $\Delta z=\d$. That is, we have only the discrete
values $z_m=m\d$. We will consider values of
$\k\d$ which range from 
0.002 to 0.128. To give a feeling for the
lattice we plot in Fig. \ref{latticefig} the $(t,x)$ coordinates of 
lattice point worldlines at intervals of 50 lattice points. 
Since $dx=a(z,t)dz$ at
constant $t$, the separation $\Delta x$ of the lattice points
in Fig. \ref{latticefig} gives a direct representation of the scale factor.
The proper lattice spacing is given by $a(z,t)\d$ which, according to 
(\ref{aH}), grows approximately linearly with time at the horizon. 

Note that there is a region where the 
lattice spacing becomes very small. This happens because we chose 
the spacing to be uniform at $t=0$. If the points did not bunch up 
outside the black hole before
$t=0$, they could not wind up equally spaced at $t=0$, since they are all on 
unit energy free-fall trajectories. (The line element is symmetric
under $(t,x)\rightarrow (-t,-x)$,  so the points bunch up also
outside the white hole horizon after $t=0$.)
 
\begin{figure}[hbt]
\centerline{
\psfig{figure=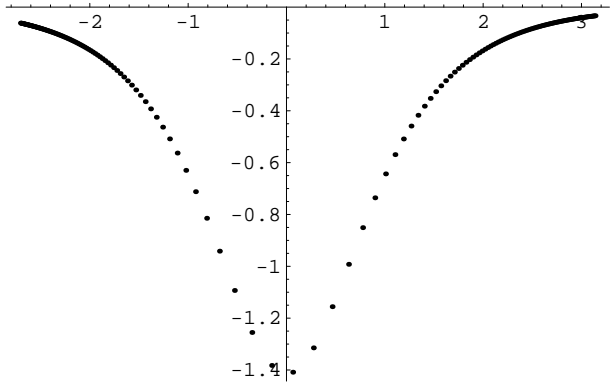,angle=0,height=5cm}}
\vskip 2mm
\caption{\small $v(x)$ for every lattice point at $t=27$, for the 
case  
$v_{\rm max}=\protect\sqrt{2}$  
and $\d=0.004$.} 
\label{vlattice}
\end{figure}

To get a better idea of how sparse the lattice is near the
horizon at the turning point of a typical wavepacket, we plot in
Fig. \ref{vlattice} the velocity function $v$ (\ref{v}) for {\it every} lattice
point near the horizon at time $t=27$, for the case $\vm=\sqrt{2}$ and 
$\k\d=0.004$. In this case there are only 6 points in the 
region between $v(x)=0.5$ and the horizon,
and the proper spacing at the horizon is $\sim \sqrt{2}\k\d t\sim 0.15/\k$.

\subsection{Scalar field}

We consider a massless, minimally coupled 
scalar field on the lattice. In the continuum the action
would be 
\begin{eqnarray}
S_{\rm cont}&=&\frac{1}{2}\int d^2x \sqrt{-g}g^{\mu\nu}
\partial_\mu\phi\partial_\nu\phi\nonumber\\
&=&\int dt\, dz\, \left((a(z,t)
(\partial_t\phi)^2-\frac{1}{a(z,t)}(\partial_z\phi)^2\right).
\label{Scont}
\end{eqnarray}
where we substituted the metric (\ref{ds2z}) in the second line.
On the lattice we adopt the same action
except that the partial derivative 
$\partial_z\phi$ is replaced by the finite difference
$D\phi_m:=(\phi_{m+1}-\phi_m)/\d$ and $a_m(t)$ is  
replaced by its average over the relevant lattice sites:
\beq
S = \frac{1}{2} \int dt \sum_m \, \left( a_m(t) 
(\partial_t \phi_m(t))^2
-  \frac{2(D \phi_{m}(t))^2}{a_{m+1}(t)+a_m(t)} 
 \right)
\label{Slat}
\end{equation}
Varying the action (\ref{Slat}) 
gives the equation of motion for $\phi_m(t)$,
\beq
\partial_t ( a_m(t) \partial_t \phi_m(t))  
- D \left(  \frac{2D \phi_{m-1}(t)}{a_{m+1}(t)+a_m(t)} \right) =0.
\label{deom}
\eeq
There is a conserved (not positive definite) 
inner product between complex solutions:
\beq
(\psi,\phi)= i \sum_m a_m(t) (\psi^{*}_m(t) \partial_t \phi_m(t)
- \phi_m(t) \partial_t \psi^{*}_m(t)).
\label{ip}
\eeq

In two dimensions the continuum action (\ref{Scont})
is conformally invariant, and all metrics are
conformally flat, hence there is no wave scattering. The discretization
breaks this conformal invariance, however there is still essentially
no scattering on the lattice providing the lattice spacing is much
smaller than the radius of curvature. The reason is that 
modes with wavelength long compared to the lattice spacing 
cannot tell they are not in the continuum, while modes with 
wavelength comparable to the lattice spacing 
cannot tell that the spacetime is curved. Thus, when we compute
the Hawking occupation numbers, there is no ``greybody factor".

\subsection{Quantization and the Hawking effect}

The field is quantized as a collection of 
self-adjoint operators $\hat\phi_m(t)$
satisfying the equation of motion (\ref{deom}) and
canonical commutation relations. For each complex solution
to the equations of motion we define the 
operator $a(f)=(f,\hat{\phi})$ using the inner product (\ref{ip}).
The commutation relations are equivalent to the relations
\beq
[a(f),a^\dagger(g)]=(f,g)
\label{ccr}
\eeq
for all $f$ and $g$.

For a positive norm solution $p$, $a(p)$ acts like
a lowering operator, while for a negative norm
solution $n$, the conjugate $n^*$ has positive norm, so
$a(n)=-a^\dagger(n^*) $ acts like a raising operator.
The Hilbert space of ingoing modes is just the Fock space
built from solutions with positive frequency with
respect to $t$ (hence positive norm), and similarly for the outgoing
Hilbert space.
We assume the boundary condition that the ingoing positive frequency
field modes are all in their ground state, 
\beq 
a(p_{in})|0\ra=0.
\label{invac}
\eeq
 
The occupation number $\la 0| a^\dagger(p)a(p)|0\ra$ 
of an outgoing normalized positive frequency
wavepacket $p$ in the in-vacuum $|0\ra$ is 
is just minus the norm of the negative frequency
part of the ingoing wavepacket that gives rise to $p$. To see this,
suppose the ingoing wavepacket $q_+ + q_-$ evolves to 
$p$ according to the field equation (\ref{deom}), 
where $q_+$ has positive 
frequency and $q_-$ has negative frequency.
Then $a(p)=a(q_+)+a(q_-)=a(q_+)-a^\dagger(q_-^*)$,
hence
\beq
\la 0|a^\dagger(p)a(p)|0\ra= -(q_-,q_-),
\label{occno}
\eeq
where we used the in-vacuum condition (\ref{invac})
and the commutation relations (\ref{ccr}). If $p$ is
not normalized then to obtain the occupation number we 
must divide by $(p,p)$.
Hawking's result\cite{Hawk75} 
in standard field theory yields, for 
a wavepacket $p=\int d\o \, c_\o \exp(-i\o(k) t+ikz)$, the 
thermal occupation number  
\beq
N_{\rm Hawking}=\int d\o \, \frac{\o|c_\o|^2}{e^{\o/T_H}-1},
\label{NHawking}
\eeq
where $T_H=\kappa/2\pi$ is the Hawking temperature.

\section{WKB ANALYSIS}
\label{wkb}
\begin{figure}[hbt]
\centerline{
\psfig{figure=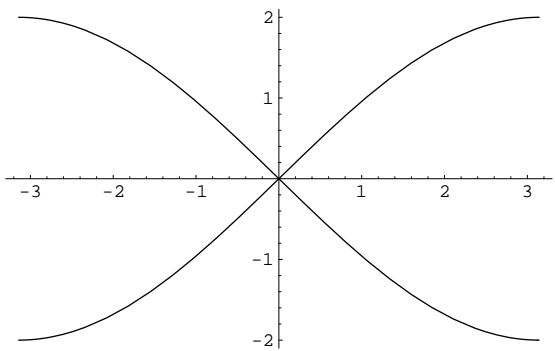,angle=0,height=4.5cm}}
\vskip 3mm
\caption{\small Dispersion relation $\o\d=\pm 2\sin(k\d/2)$ 
plotted vs. $k\d$.} 
\label{disprel}
\end{figure}

The motion of wavepackets in the
falling lattice model can be studied using the WKB
approximation\cite{CJlattice}. This amounts to using
Hamilton's equations for the position $z$ and momentum $k$
with a Hamiltonian determined by the dispersion relation 
\beq
H(z,k,t)=\o=\pm\frac{2/\d}{a(z,t)}\sin(k\d/2)
\label{H}
\eeq
which is obtained by inserting a mode of the 
form $\exp(-i\o t + ikz)$ into the field equation
and keeping only the terms with the highest derivatives 
or finite differences of the field\cite{HamBMPS}.  
A plot of this dispersion relation is shown in Fig. \ref{disprel}
for $a(z,t)=1$.
Wavevectors differing by $2\pi/\d$ are equivalent, so only the
range $(-\pi/\d,\pi/\d)$ (the ``Brillouin zone") is plotted. 
 
The frequency $\o$ is 
not conserved, even when $k\d\ll1$, simply
because it is not the Killing frequency. Using 
(\ref{kvz}) we see that in the continuum the Killing
frequency would be given by 
\beq
\o_K=\o+\frac{v(x(z,t))}{a(z,t)}k,
\label{wK}
\eeq
with which the dispersion relation can be expressed as
\beq
a(z,t)\o_K  - v(x) k=(2/\d) \sin(k\d/2).
\label{wKdisprel}
\eeq
Note however that this relation is only valid when
$k\d$ is not too large, otherwise the 
fact that the Killing frequency involves
a partial derivative in a direction that is not along a
lattice point worldline renders the relation meaningless.

It was shown in \cite{CJlattice} that the WKB trajectory of an ingoing
positive frequency wavepacket bounces off the horizon and comes 
back out provided the
ingoing wavevector is greater in magnitude than some critical 
value $|k_c|$. 
(This critical value depends on the time and place from which the
packet is launched.) The wavevector evolves from negative values
to positive values by decreasing until it
is less than $-\pi/\d$, at which point the group velocity 
(in the free fall frame) changes sign and the wavevector is
equivalent to $+\pi/\d$. From there it continues to decrease,
winding up small and positive. This reversal of group velocity
by a monotonic change of the wavevector on a lattice is analogous
to the Bloch oscillation of an electron in a crystal acted on
by a uniform electric field. 

Thus the interval $(-\pi/\d,-|k_c|)$ is mapped onto 
the interval $(0,\pi/\d)$.
The map is onto since every outgoing wavevector arises from some ingoing
wavevector. This dynamical stretching of the $k$-space interval of length 
$\pi/\d-|k_c|$ to one of length $\pi/\d$ is not problematic in a 
particle phase flow, but if we think about the wavepackets that are
being approximated by this WKB analysis it is disturbing 
since there really are more outgoing modes than ingoing modes which 
produced them. It seems the wave evolution is somehow not conserving
the number of wave degrees of freedom. 

The resolution of this puzzle is the essence of the Hawking effect.
The wave evolution mixes positive and negative frequency modes,
so a purely positive frequency outgoing wavepacket arises from a mixture
of positive and negative frequency ingoing wavepackets.  
WKB evolution breaks down at the turing point, hence it misses this
mode mixing.  
 
\begin{figure}[hbt]
\centerline{
\psfig{figure=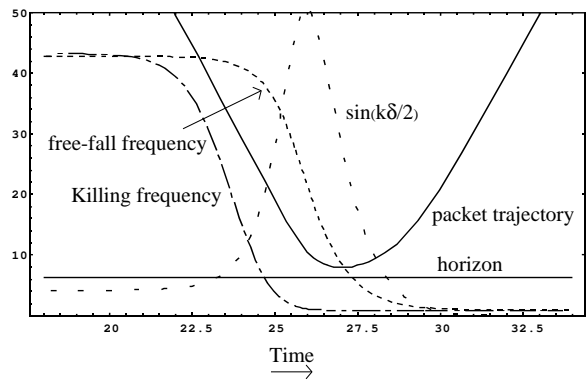,angle=0,height=5cm}}
\caption{\small WKB evolution.} 
\label{wkbgraph}
\end{figure}

An example of a positive frequency WKB trajectory is shown in 
Fig. \ref{wkbgraph}. 
The graph shows several different quantities as functions 
of $t$ (in units with $\k=1$): the static position coordinate 
of the wavepacket $x$
and the horizon $x_H$ (solid lines), the free-fall frequency $\o$ (\ref{H})
(dashed line), 
the Killing frequency $\o_K$ (\ref{wK}) (dot-dashed line),
and $50\sin(k\d/2)$ (sparsely-dashed line). 
The positions are scaled up
by a factor of 10 so they can be 
seen on the same graph. In this example the lattice spacing is 
$\d=0.004$, $\vm=\sqrt{2}$, and the trajectory has $z(34)=6$ and $k(34)=1$.

Notice first the behavior of $\sin(k\d/2)$. Backwards in 
time this starts out very small, rises to the maximum, and comes
back down but not to zero. The group velocity in the frame falling
with the lattice vanishes at the top, when $k\d/2=\pi/2$,
corresponding to a wavelength equal to two lattice units.
This rise and fall happens via a monotonic change of $k$, and is the 
Bloch oscillation mentioned above.

The wavepacket that ends up with $\o=\o_K=1$ at late times
starts out at early times with $\o=\o_K\sim 43$. 
The maximum possible lattice frequency is $2/\d=500$ in this case, 
so the ingoing frequency is certainly ``high".
Whereas the free-fall frequency continues to
redshift (forward in time) as the wavepacket climbs away from 
the horizon, the  redshifting of the Killing frequency is 
essentially complete by the time the wavepacket reaches the
turning point.
Thus, past the turning point, each Killing frequency component
propagates autonomously, more or less as it would in the 
continuum. 
As described in the Introduction, this, together
with the fact that the evolution of $\o$ at the lattice scale
is adiabatic, is the reason why the usual Hawking effect is reproduced.

\section{LATTICE WAVE EQUATION ANALYSIS}
\label{numerical}
In this section we discuss the results of our numerical calculations.
The basic method is to take an outgoing positive
frequency wavepacket and evolve it backwards in time. After it has 
``bounced" from the horizon and propagated back into the flat
region of the spacetime we decompose it into positive and
negative frequency parts. Since it is a purely left moving wavepacket at 
this stage the negative frequency part is just the part composed of positive
wavevectors. The occupation
number (\ref{occno}) of the original outgoing packet is then
given by minus the norm of this negative frequency part 
divided by the total norm. We typically computed the total norm 
in position space using (\ref{ip}). Since the negative frequency
part $\phi^{(-)}$ 
was projected out in the wavevector space we used an alternate
expression for its norm, $(\phi^{(-)},\phi^{(-)})=
-2\sum_{k>0}|\o(k)||\widetilde{\phi}^{(-)}(k)|^2$, where
the frequency $\o(k)$ is given by (\ref{H}).
 
\subsection{Numerical issues}
Our system of equations (\ref{deom}) is a coupled set of ordinary
differential equations, which is to be solved in the limit that
time is continuous but the spatial lattice is fixed. In order to satisfy
the Courant stability criterion the time step $\Delta t$ must satisfy
$\Delta t<a(z,t)\delta$. (In this 1+1 dimensional setting the
Courant condition coincides with the condition that causality
be respected by the differencing scheme.) 
In the asymptotic region $a(z,t)=1$ so the stability condition there is 
$\Delta t< \delta$. For a fixed $\Delta t/\delta$ 
the general condition would be violated wherever $a(z,t)<\Delta t/\delta$.
Since this only happens deep behind the horizon (for $t>0$)  
we just modified $a(z,t)$ inside the black hole so that
it is never less than $\Delta t/\delta$. 
We used a time discretization
scheme (written out in the Appendix)
with errors of order $(\Delta t)^2$, and found that
$\Delta t = 0.4\delta$ was adequately small to obtain very good accuracy. 

\subsection{Wavepackets and parameters}

If the Killing frequency were conserved as it is in the
dispersive continuum models the entire Hawking spectrum
could be probed just by propagating one wavepacket and
decomposing it into is Fourier components, each of which 
would pick up a negative frequency part appropriate for the
corresponding frequency. This is how Unruh checked the spectrum
in his model\cite{Unruh95}. Since the Killing frequency is not
conserved on our lattice we cannot avail ourselves of this 
option.\footnote{We studied the possibility that the WKB evolution could
be used to establish a mapping between in and out frequencies,
thus allowing us to check the entire spectrum of particle 
creation just by propagating a single wavepacket. This turns
out to be a flawed idea however since the modes are mixed by the
evolution and, moreover, such a map misses the role of the
negative frequency piece.} Thus instead we compute the occupation
numbers for a sequence of wavepackets $\widetilde{\phi}_s(k)$ 
with different profiles:
\beq
\widetilde{\phi}_s(k)=k\, \exp[-(k-0.05 s)^2],\qquad s=1,\dots,9
\label{wps}
\eeq
(in units with $\k=1$, as usual). The form of these wavepackets was
dictated by the need to have them well enough localized to be
contained in the flat region of the spacetime (so we could construct
the corresponding positive frequency initial data) while at the same
time containing enough power at low frequencies of order $\k$ to have
a measurable Hawking occupation number. To shift the wavepacket to
the desired starting position the Fourier transform (\ref{wps}) 
was multiplied by a phase factor $\exp(-ikz_{\rm initial})$.
The positive frequency
initial data was constructed by evolving each Fourier component
one time step $\Delta t$ by multiplying 
with the phase factor $\exp(-i\o(k)\Delta t)$ and Fourier transforming
the result back into position space.

The time dependence of the lattice (illustrated in Fig. \ref{latticefig})
means that the results will not be independent of the ``launch time" of the
wavepacket from a given location. If we start our backwards evolution
too far in the future then when the wavepacket nears the horizon the lattice
points will be pathologically sparse. On the other hand if we launch 
backwards at too early a time then the wavepacket can become partially
trapped in the region where the lattice spacing gets very small. 
This happens because a sufficiently high frequency wave is turned
around when it tries to propagate into a region of larger lattice
spacing. We encountered this ``channeling" phenomenon in our simulations,
and avoided it just by launching later. The runs reported here were 
all done with wavepackets that reached the horizon around
$t=27$, and were launched from far enough away to be contained in the
flat region of the spacetime. Typically, the launch time was around
$t=56$ when the wavepacket was centered at $x=28$. 

All of the calculations reported here were done using 
the line element (\ref{ds2}), with $\vm$ (\ref{v}) ranging from
$\sqrt{2}$ to $30$.

\subsection{Results}

\subsubsection{Generalities}
The behavior of a typical wavepacket throughout
the process of bouncing off the horizon is illustrated
in Fig. \ref{tsequence}. 
The real part of the
wavepacket is plotted vs. the static coordinate $x$ at several
different times. Following backwards in time,
the wave starts to squeeze up against the
horizon and then a trailing dip freezes and develops oscillations that grow
until they balloon out, forming into a compact high frequency wavepacket
that propagates neatly away from the horizon backwards in time. 

This ingoing wavepacket contains both positive and  
negative frequency components, in just the right combination 
to produce only an 
outgoing wave when sent in towards the black hole, 
with no wave propagating across the horizon. To illustrate this
we have taken one of these ingoing wavepackets, decomposed it into
its positive and negative frequency parts, and sent them separately
back toward the horizon forward in time (Fig. \ref{posnegw}). 
The negative frequency part (which is the smaller of the two since the
surface gravity is fairly low compared to the typical frequencies
in the outgoing wavepacket), mostly crosses the horizon, with just a 
bit bouncing back out. The
positive frequency part mostly bounces, with just a little
bit going into the black hole. 

\subsubsection{Spectrum}
The ``observed" occupation numbers for the
wavepackets (\ref{wps}) all
agree well with the integrated thermal predictions (\ref{NHawking}) 
of the Hawking effect provided the lattice spacing is
not too large.  
\begin{table}
\caption{Comparison of the black hole radiation on the lattice to
the thermal Hawking occupation numbers (\ref{NHawking})
for the wavepackets (\ref{wps}) for the case 
$\vm=\protect\sqrt{2}$ and $\k\d=0.002$.}
\begin{tabular}
{cccc} 
s  &  Thermal  & Lattice & Rel. Diff.  \\ \hline
1&0.01563&0.01557&0.0038\\ 
2&0.01409&0.01404&0.0035\\ 
3&0.01266&0.01262&0.0034\\ 
4&0.01135&0.01130&0.0041\\ 
5&0.01014&0.01009&0.0045\\ 
6&0.00903&0.00899&0.0039\\ 
7&0.00801&0.00797&0.0051\\ 
8&0.00709&0.00705&0.0052\\ 
9&0.00625&0.00622&0.0046\\ 
\end{tabular}
\label{table}
\end{table}
A sample of the comparisons for $\k\d=0.002$
is listed in Table \ref{table}. The typical relative difference
in this case is of order half a percent. 

\subsubsection{Lattice dependence}
To study the dependence on the lattice we ran the $s=1$
wavepacket backward at several different values of the lattice 
spacing and for several values of $\vm$ (\ref{v}) from $\sqrt{2}$ to 30.
The change of $\vm$ produces only a rather small change in 
the shape of the function $v(x)$ outside the horizon, however 
it produces a shift of the position of the horizon (\ref{xH}). 
For large $\vm$ the horizon is given approximately by
$x_H\simeq\ln(2\vm)$. The horizon position ranges from about 0.6
for $\vm=\sqrt{2}$ to about 4 for $\vm=30$.  
Thus a wavepacket launched backward in time from a given 
position at a given time will reach the horizon
later in time for larger $\vm$, so that the proper lattice spacing 
at the horizon will be larger, though not by a very large amount
for the wavepackets considered here.

\begin{figure}[htb]
\centerline{
\psfig{figure=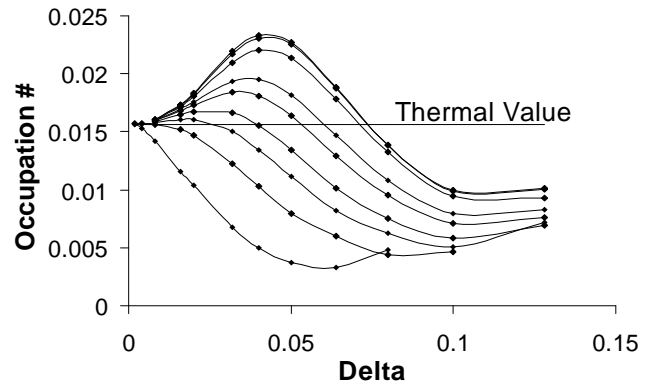,angle=0,height=5cm}}
\vskip 3mm
\caption{\small Occupation numbers for $s=1$ wavepacket for various
values of $\d$ and $\vm$. Higher occupation numbers correspond
to larger values of $\vm$.} 
\label{chart}
\end{figure}

In Fig. \ref{chart} we plot the occupation numbers for 
lattice spacings $\d$ ranging from 0.002 to 0.128,
and for several values of $\vm$ (\ref{v}),  
$\vm=\sqrt{2},2,2.5,3,4,5,10,20,30$.
In all cases the late time wavepacket was launched from 
the same position and time. All wavepackets had the same 
envelope for their Fourier transform, however for different 
lattice spacings the wavepacket is necessarily different.
For all values of $\vm$, the occupation number converges to the
Hawking prediction as the lattice spacing $\d$ decreases. 

The pattern of deviations from the thermal prediction appears
to depend quite strongly on the value of $\vm$, a result which
we do not understand at present. The best we can do is to list
various effects which would contribute to the deviations: 
(i) the WKB turning point
is moving away from the horizon as $\d$ grows, so one might expect that the 
effective surface gravity $v'(x_{\rm t.p.})$ and hence the effective 
Hawking temperature felt by the wavepacket
would decrease;  
(ii) at larger $\vm$ the proper lattice spacing at the horizon is
larger at the time this particular wavepacket reaches the horizon;
(iii) the lattice points are rather sparse near the horizon for the
larger values of $\d$, and their precise positions will be significantly
different for different values of $\vm$.
Effect (i) would lead to a decrease in the occupation number, while
one might expect that effect (ii) would lead to a relative increase
due to the extra particle creation associated with time dependence
of the lattice. Perhaps what we are seeing is just  
a delicate balance between these two effects.

\section{DISCUSSION}
\label{discussion}
We have shown that linear field theory on a falling lattice 
reproduces the continuum Hawking effect with high fidelity,
thus verifying earlier expections\cite{CJlattice}. 
For the smallest lattice spacing we studied, $\d=0.002/\k$,
the deviations from the thermal particle creation in
wavepackets composed of frequencies $\o\lsim O(\k)$ were of order
half a percent. This is remarkable, particularly since the {\it proper}
lattice spacing encountered by the wavepacket at the horizon
was $\sim0.08/\k$. When $\k\d$ is pushed to higher 
values eventually we see significant deviations from the Hawking
prediction.

There are several directions in which this work could be 
developed. One idea is to include self-interactions for the
quantum field theory. There are general 
arguments\cite{GibbonsPerry,UnruhWeiss} that
the thermal Hawking effect holds also for interacting fields.
There are also
perturbative calculations\cite{UnruhLeahy} and 
calculations exploiting
exact conformal invariance of certain 1+1 dimensional 
field theories\cite{BirrellDavies}. Also 
the ``hadronization" of Hawking radiation from primordial
black holes has been analyzed by matching to high energy collision
data\cite{hadronization}, but the
physics of how the thermal Hawking radiation is ``dressed" by
interactions as it redshifts away from the horizon has not 
been studied.
In a 1+1 dimensional model it should not
be prohibitively difficult to numerically study Hawking radiation 
of interacting fields, now that we see it can be done on
a lattice.  

The falling lattice model has provided a satisfactory
mechanism---the Bloch oscillation---for how to get an outgoing
mode from an ingoing mode in a stationary background, but there
is a serious flaw in the picture: the lattice is constantly
expanding. In the fluid model by contrast the lattice of atoms maintains
a uniform average density. In a fundamental theory we might also
expect the scale of graininess of spacetime to remain fixed 
at, say, the Planck scale or the string scale (since presumably
the graininess would {\it define} this scale) rather than 
expand. Can the falling lattice model be improved to share this
feature? 

A fluid maintains uniform density in an 
inhomogeneous flow by compressing in some directions and 
expanding in others, which requires at least two dimensions
to be possible. At the atomic level such a volume-preserving
flow involves erratic motions of individual atoms. One possible
improvement of the lattice model is to make a lattice
that mimics this sort of volume preserving flow. It is not 
clear whether the motions of the lattice points can be slow
enough to be adiabatic on the timescale of the high frequency
lattice modes. If they cannot, then the time dependence of 
this erratic lattice background will excite the quantum vacuum.

In a fluid the lattice is a part of the system, not just a fixed
background. The average flow could be adiabatic for the fully coupled 
ground state of the system but not for the field theory of the perturbations
on the ``background lattice". Similar comments apply in quantum
gravity: surely if in-out mode conversion is at play the incoming 
high frequency modes are strongly coupled to the quantum gravitational
vacuum. Ideally, therefore, we should try to find a model in which
the background is not decoupled from the perturbations.

As a first step in this direction one could study a
one dimensional model in which the lattice points are non-relativistic
point masses, coupled to each other by nearest neighbor interactions, 
and ``falling" or propagating in a background potential (with
or without periodic boundary conditions).
The perturbations of
such a lattice are the phonon field, and the back reaction to the 
Hawking radiation is included
(although the background potential is fixed). 
In a model like this one could presumably follow in detail the 
nonlinear origin of the outgoing modes and the transfer of 
energy from the mean flow to the thermal radiation.

\section*{Acknowledgements}
We are grateful 
to Matt Choptuik and David Garfinkle for advice on numerical matters
and to Steve Corley for many helpful discussions. 
This work was supported in part by the National Science Foundation
under grants No. PHY98-00967 
at the University of Maryland and PHY94-07194 at the Institute for 
Theoretical Physics.  

\newpage

\appendix 
\onecolumn
\section*{Finite differencing scheme}
 
The equation of motion (\ref{deom}) for $\phi_m(t)$ is
\beq
\partial_t \Bigl( a_m(t) \partial_t \phi_m(t)\Bigr)  
= D \left(  \frac{2D \phi_{m-1}(t)}{a_{m+1}(t)+a_m(t)}\right).
\label{deom2}
\eeq
We used a time discretization
scheme for the left hand side arising from the replacement
of $\dot{f}$ by 
\widetext
\beq
\Bigl[f(t+\Delta t/2)-f(t-\Delta t/2)\Bigr]/\Delta t= 
\dot{f} + (1/24) {f}^{(3)}(\Delta t)^2 + O((\Delta t)^4), 
\eeq
which yields
\beq
\partial_t ( a_m(t) \partial_t \phi_m(t))\rightarrow
\Bigl[a(n+1/2)\Bigl(\phi(n+1)-\phi(n)\Bigr)-
a(n-1/2)\Bigl(\phi(n)-\phi(n-1)\Bigr)\Bigr]/
(\Delta t)^2,
\eeq
where the spatial index is omitted and the argument $n$ stands
for $n\Delta t$. We could have used this replacement, since $a(z,t)$
is a prescribed function that can be calculated at any $t$, but
for some reason we substituted $a(n\pm 1/2)$ by the average
$\Bigl(a(n)+a(n\pm1)\Bigr)/2$. This yields the approximation
\begin{eqnarray}
 \Bigl[\Bigl(a(n)+a(n+1)\Bigr)\Bigl(\phi(n+1)-\phi(n)\Bigr)- 
 \Bigl(a(n)+a(n-1)\Bigr)\Bigl(\phi(n)-\phi(n-1)\Bigr)\Bigr]/
2(\Delta t)^2\label{lhs}\\
=\partial_t ( a  \partial_t \phi )(n) +
\Bigl(\frac{1}{12} {a} {\phi}^{(4)} +
\frac{1}{6}\dot{a} {\phi}^{(3)} +
\frac{1}{4}\ddot{a}\ddot{\phi} +
\frac{1}{4} {a}^{(3)}\dot{\phi}\Bigr)(n)(\Delta t)^2 + O((\Delta t)^4).
\end{eqnarray}
With (\ref{lhs}) in place of the left hand side of (\ref{deom2})
we solve for $\phi(n-1)$ in terms of $\phi(n)$ and 
$\phi(n+1)$ to obtain an equation for evolving $\phi$
backwards in time by one time step given the following two
steps.

\begin{figure}[t]
\centerline{
\psfig{figure=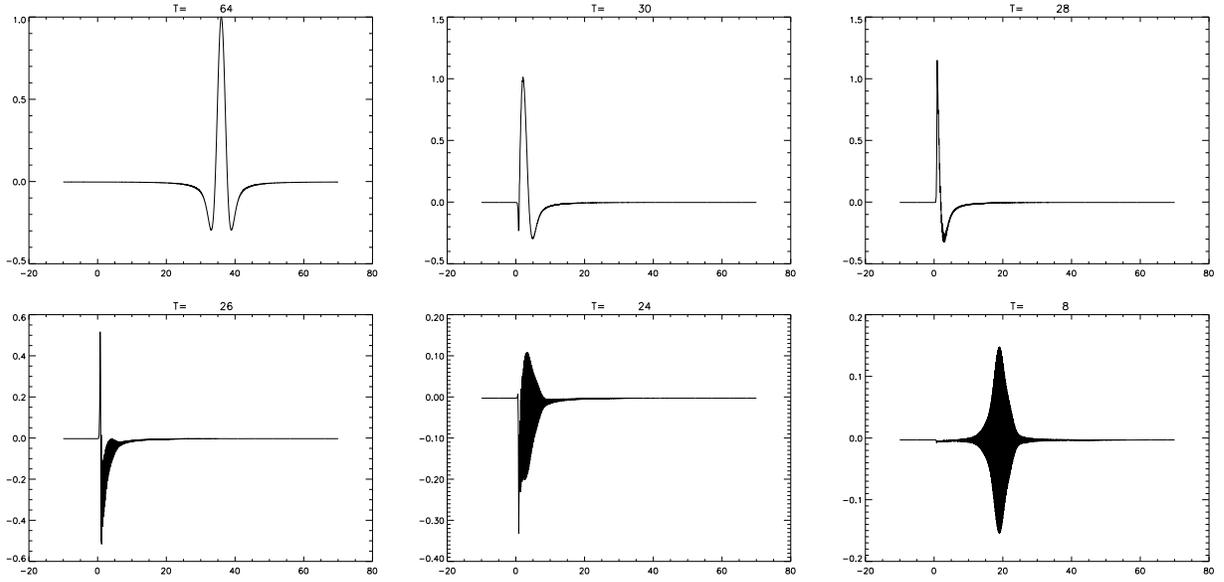,angle=0,height=22cm}}
\vskip -10cm
\caption{\small A typical wavepacket evolution. Here 
$\vm=\protect\sqrt{2}$,
$\k\d=0.004$, and the 
wavepacket (\ref{wps}) has $s=1$. The ocsillations of
the incoming wavepacket are too dense to resolve in the plots.} 
\label{tsequence}
\end{figure}
 
\begin{figure}[t]
\centerline{
\psfig{figure=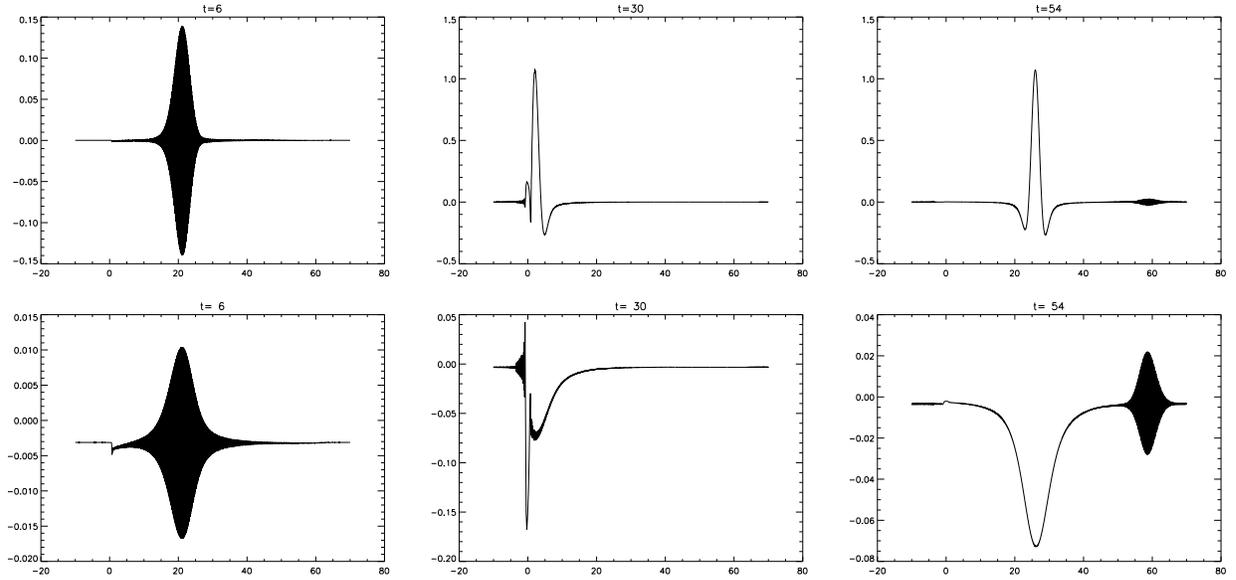,angle=0,height=22cm}}
\vskip -10cm
\caption{\small The positive (top row) and negative (bottom row) 
frequency parts of the ancestor
of an outgoing positive frequency wavepacket, 
sent back towards the black hole. Note the vertical 
scale is different in the two plots. 
The wavepackets crossing the horizon in the two cases are negatives of
each other since they must cancel to produce only the outgoing
wavepacket we started with.
The parts that go through the
horizon appear on the right due to periodic boundary
conditions. The high frequency ocsillations  
are too dense to resolve in the plots.} 
\label{posnegw}
\end{figure}


\begin{thebibliography}{99}
\bibitem{HamBMPS}The Hamiltonian technique was first applied to 
dispersive models of the Hawking effect in ref. \cite{BMPS}.
\end{thebibliography}
\end{document}